# Comprehensive analog signal processing platform enabled with acoustic charge transport in two-dimensional materials

Yueyi Sun[1], Siming Liu[1], Yingjie Luo[1], Jiwei Chen[1], Yihong Sun[1], and Changjian Zhou[1*]

1. School of Microelectronics, South China University of Technology, Guangzhou 511442, P. R. China

*E-mail: zhoucj@scut.edu.cn

Two-dimensional Acoustic Charge Transport (2D-ACT) devices, which integrate two-dimensional semiconductor field-effect transistor (FET) with high-frequency surface acoustic wave (SAW) device provide a potential compact platform for the processing of analog signals in a wireless, non-contact, low-loss and real-time way. It is expected to be used in long-distance space communication and sensing. However, current investigations into 2D-ACT devices are still limited to the observation of DC acoustoelectric currents, and have yet to achieve real-time electronic signal processing capabilities. In this paper, we have designed a hybrid acoustoelectric platform composed of two-dimensional semiconductor FET and SAW device. The platform is capable of processing DC signals, exhibiting ambipolar transport behavior. The sub-wavelength channel length of the FET within the platform allows for the real-time observation of carrier distribution at a microscopic scale in conjunction with the SAW potential, and facilitating the reproduction and intensity regulation of AC signals. By adjusting the relative phase and intensity ratio of two counter-propagating SAWs, the platform also enables the addition and subtraction of AC signals.

The realization of real-time analog signal processors, such as transversal filters, correlators, and convolvers, relies on successfully achieving a linear function of time delay, along with the capability to process analog signals. Acoustic Charge Transport

(ACT) devices employ surface acoustic wave (SAW) to create a moving potential wave in semiconductor materials. This wave captures, manipulates, and transports charge carriers in a specific direction. SAWs have a wavelength that is five orders of magnitude shorter than electromagnetic waves of the same frequency. This feature shows great potential for the device's miniaturization. The combination of semiconductors and piezoelectric acoustic materials enables the development of miniaturized RF acoustic-electric systems with low power consumption[1,2]. ACT devices also facilitate non-contact measurements, which avoid the formation of Schottky barriers, band movement, and alteration of carrier mobility[3,4].

As early as the 1990s, gallium arsenide (GaAs) was utilized to fabricate buried-channel ACT devices[5]. Later, SAWs were employed to control individual electron and photon in GaAs quantum wells, which were then applied to quantum computing[6-9]. However, these GaAs ACT devices exhibit certain limitations: the lightly doped epitaxial layers used to form the channel are excessively thick (5 μm), making monolithic integration with field-effect transistor (FET) challenging. Furthermore, charge transport occurs deep within the depleted epitaxial layer, where the SAW potential significantly diminishes, thereby decreasing the transport efficiency of the ACT device and increasing its energy loss. Also, the implementation of back-gate control at the interface between the epitaxial layer and the substrate poses significant difficulties. Two-dimensional (2D) materials with atomic-scale thickness are promising candidates for developing next-generation ACT devices with multiple functions due to their widely tunable physical properties by various stimuli. Researchers have utilized graphene to fabricate ACT devices with high carrier concentrations[10-13], however, it is well-known that the zero bandgap of graphene is unsuitable for applications requiring high on-off ratios. Subsequently, attention has shifted to semiconducting transition metal dichalcogenides (TMDCs)[14-18], BP[19] and $SnS_2$[20] which possess tunable band gaps and excellent optoelectronic properties. However, current researches are limited to the study of direct current (DC) acoustoelectric current, and there are no mature technological ways to regulate the switching behavior of ACT devices. To fully integrate ACT devices into real-time electronic signal processing systems and perform specific operations, ACT devices must possess the capability to process alternating current (AC) signals.

In this paper, we have constructed a hybrid acoustoelectric platform by integrating $WSe_2$ FET and SAW devices. The functionalities of this platform are depicted in Fig.1a. By altering the gate voltage, the platform is able to modulate both the carrier polarity and concentration, which resulted in observable positive or negative DC acoustoelectric currents. We design the FET channel length in the platform to be sub-wavelength, approximately half the wavelength of the SAW. This configuration enables the real-time response of carriers to the SAW acoustic potential at a microscopic scale, facilitating the reproduction and intensity regulation of AC signals. Ultimately, by using standing surface acoustic waves (SSAWs), we executed addition and subtraction operations for analog signals. This work also enriches the theoretical understanding of the interaction mechanisms between acoustoelectric charges and SAWs in 2D materials. The results provide an important insight into the coupling mechanisms between SAWs and charge carriers in 2D semiconductor systems. Our hybrid acoustoelectric platform offers a promising solution for spatial communication and sensing providing long-range, low-loss, and contactless capabilities that are essential for high-quality real-time AC signal processing.

## Integrated 2D Material Acoustoelectric Signal Processing Platform's functionality and structure

Our integrated 2D material acoustoelectric signal processing platform, with fabrication details outlined in the Methods section, comprises two primary components, as shown in Fig.1a: (i) a set of interdigital transducers (IDTs) constructed on a 128°YX-cut $LiNbO_3$ substrate for the actuation and detection of SAWs, and (ii) a $WSe_2$ FET device with a h-BN dielectric layer functioned as a gate dielectric and Gr as the back gate electrode, centered in between the two IDTs. The IDT fingers' width and spacing between adjacent IDT fingers are both 6.25 μm, corresponding to an operational frequency of approximately 160 MHz, the SEM image is shown in Fig.1b. The distance between the set of IDTs is 2.5 mm, with an acoustic wave propagation velocity of 3980 ms$^{-1}$. The RF characteristics of the IDTs were characterized using a vector network analyzer. The scattering parameters $S_{11}$ (reflection) and $S_{21}$ (transmission and insertion loss) are shown in Fig.1c. The $S_{21}$ parameter at the resonant frequency is 8 dB, indicates that the LiNbO$_3$ piezoelectric substrate is proficient in SAW generation, transmission, and detection. The FET has a channel length of 12.3 μm, the length is approximately equal to half the wavelength of SAW. Source and drain electrodes were sequentially deposited with 20 nm of Ti and 40 nm of Au. The overall optical microscope image of the FET is illustrated in Fig.1d. The inset displays that the thickness of WSe$_2$ and h-BN are 9.2 nm and 12 nm, respectively, measured by atomic force microscopy (AFM). Fig.1e presents a detailed schematic illustration that delineates the relative position and structural configuration of the IDTs and the FET. The PL and Raman test results in Fig.1f and Supplementary Fig.1 confirm that WSe$_2$ exhibits typical multilayer characteristics.

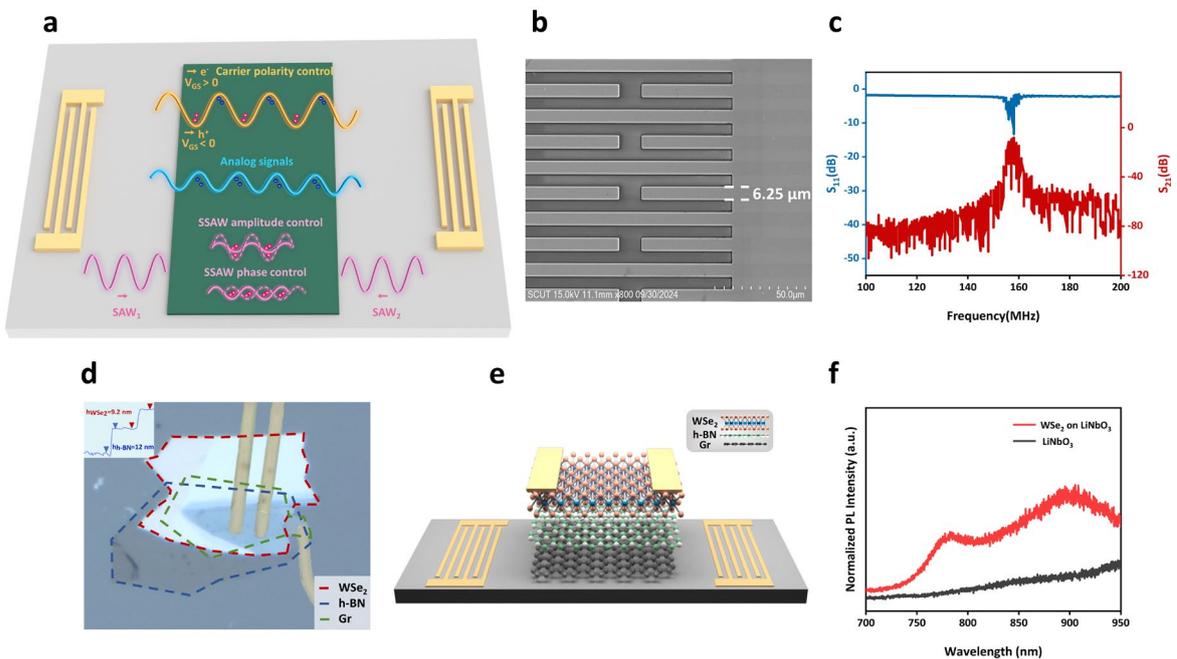

**Fig.1 | Platform's functionality, IDTs' RF characteristics, and FET structure overview. a**, Schematic of the functionalities of our acoustoelectric platform, including charge carrier bipolar regulation, DC signal and AC analog signal handling, as well as addition and subtraction operations of analog signals and reproduction of output signals using SSAW amplitude and phase modulation.

**b**, SEM images of the IDTs. **c**, $S_{21}$ and $S_{11}$ parameters measured as a function of frequency of the IDTs. **d**, The optical microscope image of WSe$_2$ FET, with the 2D material thickness measured by AFM depicted in the inset. **e**, Schematic diagram of the relative position and structure of IDTs and FET. **f**, The PL spectrum of the LiNbO$_3$ substrate and the WSe$_2$ on the LiNbO$_3$ substrate.

## Ambipolar transport characteristics of the 2D FET

First, we evaluate the 2D FET electrical properties and charge transport behavior. The dielectric constant of the h-BN gate dielectric layer is $\varepsilon_r \sim 3.9$. Due to the small thickness of h-BN ($d_{\text{h-BN}} \approx 12$ nm), the gating electric fields can reach $E = \varepsilon_r V_{GS}/d_{h-BN} = \pm 1.625 \times 10^4$ kVcm$^{-1}$ at a moderate gate bias of $V_{GS} = \pm 5$ V. We deposit Ti/Au metal electrodes to form ambipolar contact, instead of traditional high work-function Pd/Au or Pt/Au[21-26] contact metal for efficient p-type hole injection. Fig.2a illustrates the transfer characteristics demonstrating ambipolar transport behavior. When $V_{GS}$ is less than -1 V, holes become the dominant carriers. Conversely, when $V_{GS}$ is higher than 2.5 V, electrons become the dominant carriers. Due to the weak Fermi level pinning[27] at the metal/WSe$_2$ interface, the Fermi energy level will move with the change of the applied electric field. Ambipolar behavior results from the variation of Fermi level relative to metal's work function. The insulating condition of WSe$_2$ around $V_{GS} = 0$ V clearly shows the intrinsic nature of the material. The p-branch threshold voltage $V_{TH}$ of -1 V and the n-branch $V_{TH}$ of 2.5 V can be determined by analyzing the linear section of the transfer characteristics. Throughout the whole measurement range, the leakage gate current remains at the level of background noise (An order of magnitude of 10$^{-11}$ A), indicating the high quality of the h-BN dielectric. Fig.2b shows the gate voltage dependent-mobility, which is calculated by $\mu = \frac{L}{W} \times \frac{1}{C_{ox}} \times \frac{dI_{DS}}{dV_{GS}} \times \frac{1}{V_{DS}}$. In this expression, $C_{ox}$ represent the capacitance of h-BN dielectric layer, L and W are the channel length and width, respectively. The peak hole mobility is measured at 154.75 cm$^2\cdot$V$^{-1}\cdot$s$^{-1}$ with a gate voltage of -4 V, while the maximum electron mobility is observed to be 145.87 cm$^2\cdot$s$^{-1}\cdot$V$^{-1}$ at a gate voltage of 5 V. The nearly symmetrical characteristics of n-type and p-type carrier transport indicate that the Schottky barrier height is comparable for both electron and hole injection. We have observed that the electron and hole mobility measured in the experiment are significantly higher in comparison to those WSe$_2$ bipolar transistors reported in previous studies[28-31]. This enhanced mobility is due to the placement of the WSe$_2$ conductive channel over an h-BN layer. The h-BN acts as a superior insulating layer, preventing the decrease in carrier mobility caused by impurity scattering from the LiNbO$_3$ substrate[32]. Fig.2c,d show the output characteristics of WSe$_2$ FETs. Nearly linear $I_{DS}$-$V_{DS}$ behaviors are observed in both the electron and hole transport regimes.

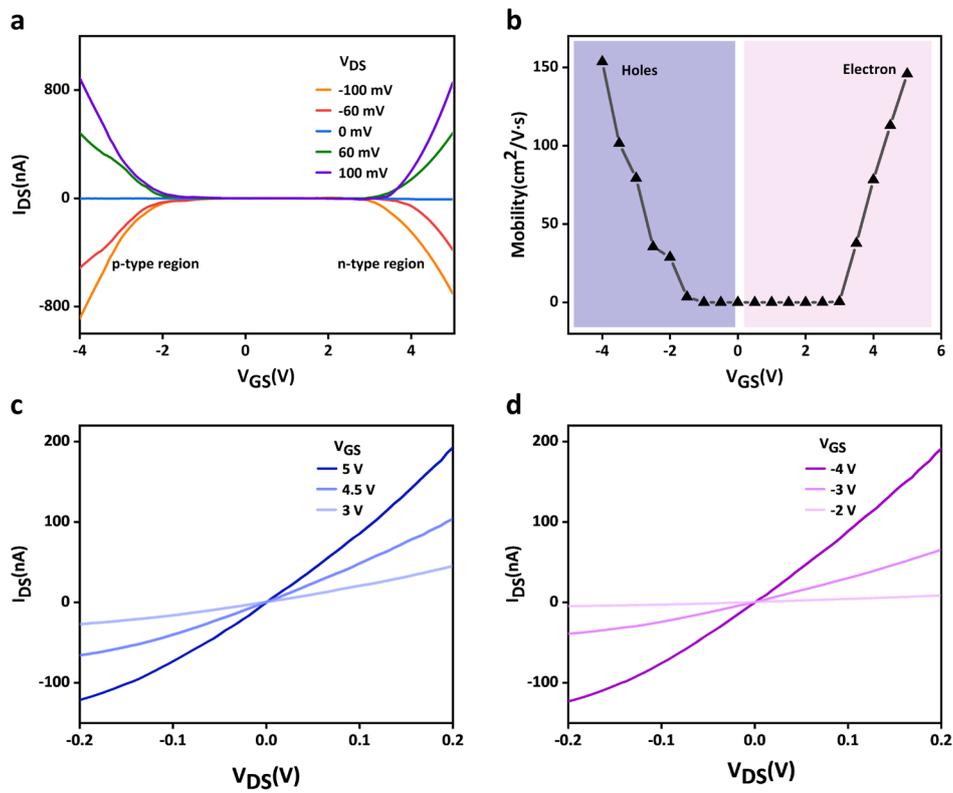

**Fig.2 | Electrical properties of the WSe$_2$/h-BN FET. a,b**, Transfer characteristic curves for different $V_{DS}$ and corresponding mobility, respectively. **c,d**, Output characteristics curves for different positive and negative gate voltages $V_{GS}$, respectively.

## Acoustoelectric effect of the ambipolar 2D FET

We further investigate the acoustoelectric effect in the WSe$_2$ after confirming its ambipolar transport characteristics. We define the propagation direction of the SAW as the positive direction, simultaneously defining the probe situated near the excitation source as Probe 1 (Drain) and the probe located further away as Probe 2 (Source), as illustrated in Fig.3a,b. Fig.3a illustrates acoustoelectric current $I_{DS}$ as a function of sinusoidal RF signal frequency under various power $P_{RF}$, when $V_{GS} = 5$ V and $V_{DS} = 0$ V. The frequency dependence of the acoustoelectric current is consistent with the transmission spectrum shown by the $S_{21}$ curve in Fig.1c. The polarity of the acoustoelectric current indicates that the majority of charge carriers are electrons. When $V_{GS} = -4$ V, the dependency of the acoustoelectric current on the frequency and $P_{RF}$ persisted, while the positive acoustoelectric current indicates that the predominant charge carriers have transitioned to holes. We then set the excitation frequency to 160 MHz and $V_{GS}$ to -6 V. Fig.3c illustrates the variations in acoustoelectric current in response to different SAW excitation power levels. Fig.3d illustrates the linear relationship between the acoustoelectric current magnitude and the excitation power of SAW at both positive and negative gate voltages. Fig.3e shows the output characteristic curves for applying a 20 dBm SAW (red curve) and without SAW (black curve) at $V_{GS} = 5$ V. The curve without SAW intersects the origin, while the SAW application shifts the curve negatively, indicating the generation of approximately -50 nA of negative acoustoelectric current. When $V_{GS} = -6$ V, holes become the majority charge carriers, causing a positive shift of the curve with SAW (Supplementary Fig.3a). Fig.3f displays the transfer curves of the FET with (red curve) and without (black curve) the 20 dBm SAW. Under negative gate voltage, SAW decreases the absolute value of $I_{DS}$, while under positive gate voltage, it increases it. This variation arises because electrons and holes move oppositely under same $V_{DS}$, resulting in the same current direction. However, under SAW excitation, their movement aligns, leading to opposite current directions. The transfer characteristic for $V_{DS} = 1$ V is shown in Supplementary Fig.3b. The potential generated by the SAW is calculated using COMSOL simulations, resulting in a sufficiently high barrier height (Supplementary Fig.2). As a result, very few electrons or holes have enough energy to cross to an adjacent well. Therefore, it can be concluded that the carriers are effectively guided by the SAW, rendering diffusion effects negligible (Supplementary Information, Note 2). Fig.3g depicts the schematic of the SAW-driven holes and electrons, leading to positive and negative acoustoelectric currents.

Then the platform's real-time response to AC analog signals is investigated by using an oscilloscope. The wiring configuration is shown on the left inset of Fig.3h. The equivalent circuit is depicted on the right. The total current between the two contacts is set to $I_{DS}$=0, and the open-circuit real-time acoustoelectric voltage (AEV) is measured. Due to the rather lower velocity of SAWs (3980 m/s) in comparison to electromagnetic waves, ACT are frequently employed in the development of delay line devices that exhibit extended delays[33,34]. At t=0 ns, the IDTs are excited, and the oscilloscope signal is observed at approximately t=280 ns. The delay time per centimeter is approximately 2.4 microseconds. The acoustoelectric current can be represented by the following formula, the detailed derivation process can be found in the supplementary materials (Supplementary Information, Note 3).:

$$I_{AE} = -\frac{1}{2}\frac{en_0c}{(1+m)^2}(\frac{q\gamma_\tau\varphi_0}{KT})^2\frac{(1-m)-u(1+m/b)}{1+(\beta_0-u)^2[(1-m)/(1+m)]^2(\omega_0/\omega)^2} \quad (1)$$

In this expression, $\gamma_\tau$ is a parameter related to the relaxation time required for a stable distribution of charge, and the potential energy generated by the SAWs can be expressed in a fluctuating form $\varphi = \varphi_0 exp(ik(x-ct))$ where $\varphi_0$ is a constant, $k$ is the wave vector, and $c$ is the velocity of the acoustic wave moving in the x-direction, $m = n_0/p_0$ is the ratio of equilibrium concentrations of electrons and holes, $q = q_n + q_p$, $q_n$ and $q_p$ are the electron and hole acoustic charges respectively, $b$ is the ratio of the electron and hole mobilities, $\beta_0$ defined as $(1-mb)/(1-m)$.

Schematic Fig.3i elucidates this phenomenon, taking electrons are the majority charge carriers as an example, at time $t_1$, the SAW carries electrons bound at the bottom of the conduction band forward and injects them into Probe 1, resulting in a pronounced sine voltage signal on the oscilloscope. At time $t_2$, as the SAW continues to move, electrons at the bottom of the conduction band are recaptured. However, due to the influences of parameter $\gamma_\tau$ and $\varphi_0$ in Formula 1, the potential of the SAW and the number of captured electrons decrease. The first parameter $\gamma_\tau$ related to the relaxation time $\tau$ needed for a stable charge distribution. Charges can respond to changes in the total potential in a time on the order of a dielectric relaxation time $\tau = \varepsilon/\sigma$[35], where

$\varepsilon$ represents the dielectric constant and $\sigma$ represents the device conductivity (calculate from Fig.2a using equation $\sigma = (I_d/V_d) \times (L/W)$), the value of $\tau$ along with the time takes for the SAW to travel pass through the channel, are on the order of ~1 ns. Therefore, the charges don't have enough time to reach an equilibrium state upon injection at Probe 2, resulting in a smaller number of charges. The second parameter is the potential energy $\varphi_0$, as the electric field traverses the heterojunction region composed of graphene, h-BN, and WSe$_2$, the electric field generated by the SAW is partially shielded, resulting in an attenuation of the SAW intensity[36,37], as shown by the simulation results in Supplementary Fig.4, which leads to a reduction in potential energy. After the recapturing, SAW begins to inject the electrons into Probe 2. However, due to the reduced number of electrons, this is reflected as a weaker sine voltage signal on the oscilloscope, with the voltage polarity reversed. The test results presented in Fig.3h also confirm the accuracy of the aforementioned analysis, as the voltage exhibits positive and negative oscillations and a difference in the absolute values of the positive ($\approx$ 160 mW) and negative voltages ($\approx$ 200 mW). From the analysis of the AC measurement results, it is discernible that the DC acoustoelectric current depicted in Fig.3a~f, measured with a Keithley 4200A-SCS source-measurement unit (SMU) originates from the integration of charge quantities sequentially injected into Probes 1 and 2 within a 5 ms time interval (the minimum sampling interval of the SMU). It is evident that the observed DC acoustoelectric current in the earlier experiments is, in fact, the integral of the number of charge carriers injected into the source and drain probes[16,38-40]. These experiments generally lack observations of real-time AC acoustoelectric current, primarily because the device channel lengths used in these studies often exceed the wavelength of the SAW by several or even dozens of times. In contrast, the device channel length in this paper lies within the sub-wavelength range, approximately half a wavelength, thereby allowing for real-time responses of AC acoustoelectric current at the microscopic scale. This discovery further enriches and enhances the theoretical understanding of the acoustoelectric effect.

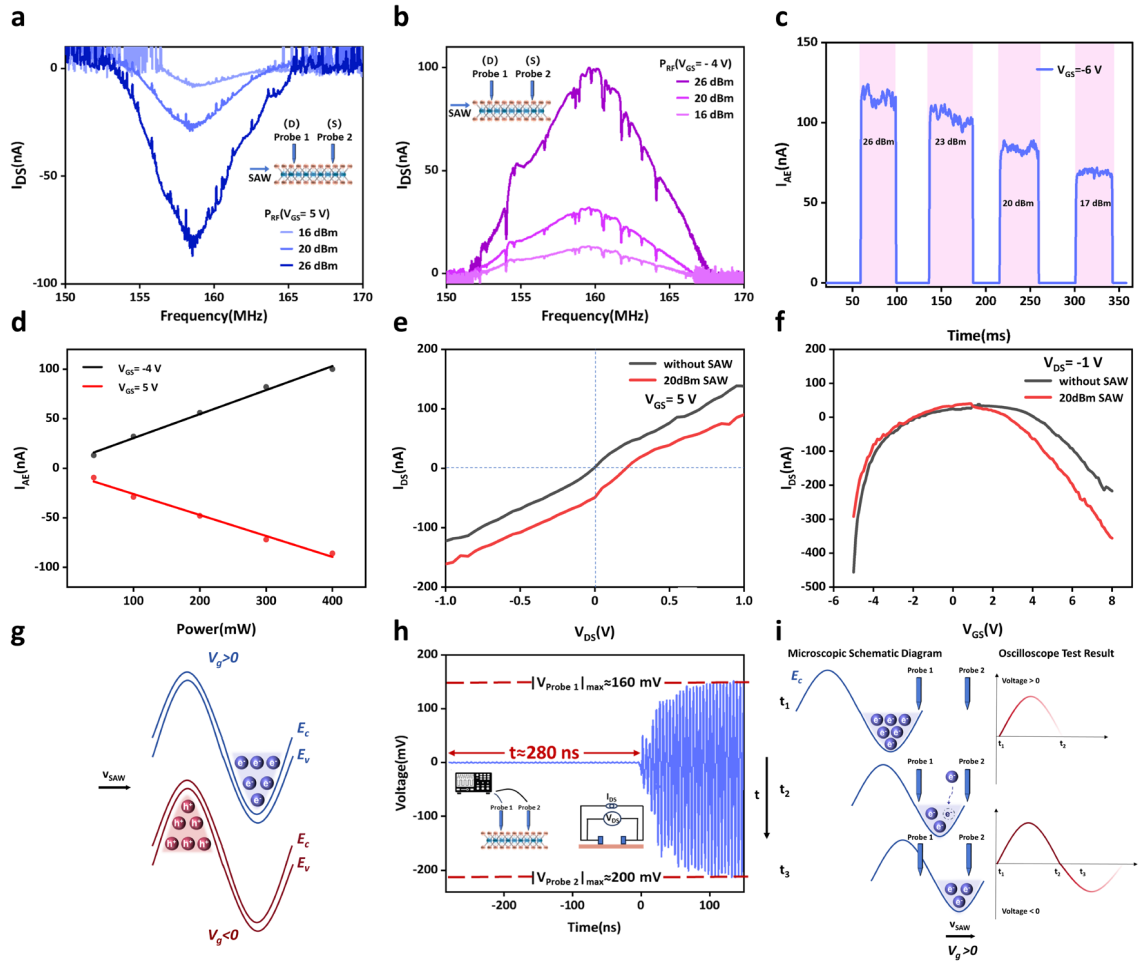

**Fig.3 | Acoustoelectric effect. a,b**, DC measurement of the dependence of acoustoelectric current on the RF signal frequency, with positive and negative gate voltage modulation, respectively. Define the propagation direction of SAW as the positive direction. **c**, Acoustoelectric current driven by the SAW with different excitation powers. **d**, The linear relationship between acoustoelectric current and SAW excitation power. **e**, Output curves of the FET when $V_{GS}$ = 5V with (red curve) and without (black curve) the SAW. **f**, Transfer curves of the FET when $I_{DS}$ was driven by the electrical field only (black curve) and the SAW as well (red curve). **g**, Schematic illustration depicting the microscopic principle of acoustoelectric current generation under RF excitation. **h**, AC measurement in the time domain, the tests were conducted using an oscilloscope (shown on the left inset), with the equivalent circuit diagram displayed on the right. **i**, A microscopic schematic illustrating the electron real-time injection mechanism during the acoustoelectric current AC testing.

## Assessment of the platform's capability in processing simple analog

**signals**

We also assessed the platform's performance in processing high-frequency analog signals. Under continuous sine wave RF excitation, with variable power and $V_{GS}$ = 5 V, Fig.4a indicates that the platform can handle signal intensities up to 26 dBm, limited by the amplifier's maximum gain, resulting in a dynamic range of approximately 18 dB for the platform with electrons as charge carriers. When $V_{GS}$ = -4 V, the dynamic range is 19 dB when holes act as the active carriers, as shown in Fig.4c. By reducing the gate control voltage from 5V to 3.2V, we examine the cutoff behavior of the platform. The output signal decreases to the noise level, indicating an on-off ratio of $10^4$ in Fig.4b. The platform's off state is reached when the gate voltage is increased from - 4V to - 1.3V, as shown in Fig.4d. The on-off ratio remains approximately $10^4$.

Additionally, we investigate the device's capacity to process pulse AC signals. Fig.2e,f respectively shows the pulse responses of the platform when a single pulse and twenty pulse excitation signals are applied to the IDTs, the responses for other pulse quantities are provided in the supplementary document (Supplementary Fig.5). It is possible to see how the voltage signal first builds up and tends toward stability, which is consistent with the frequency characteristics of the SAW response[41]. Comparing our studies to previous works, we are able to obtain a better signal-to-noise ratio when using pulsed SAWs to drive the acoustoelectric current[42]. This not only demonstrates the device's potential for processing analog signals but also simultaneously reduces the impact of thermal effects on the device.

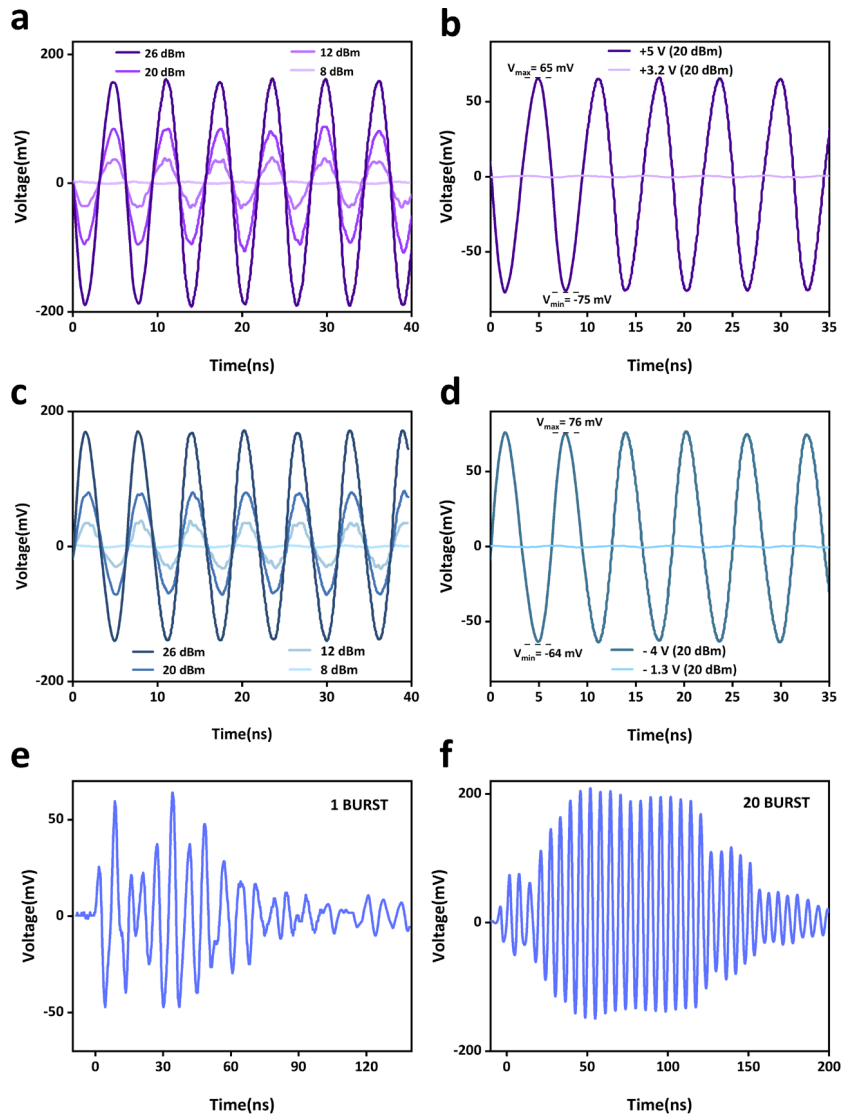

**Fig.4 | The acoustoelectric response of the platform to simple AC signals. a,c**, Variations in the voltage across the device caused by acoustoelectric current as the RF excitation power changes, measured via oscilloscope testing, with gate voltages fixed at 5 V and -4 V, respectively. **b,d**, Positive and negative gate control of the device on/off characteristic, respectively. **e,f**, Pulse responses measured under single and twenty RF pulse excitations.

## Addition and subtraction operations on analog signals

When two counter-propagating SAWs interfere, as illustrated in Fig.5a, a standing surface acoustic wave (SSAW) is formed. The position and properties of the SSAW can be controlled by adjusting the phase and amplitude of the two counter-propagating SAWs. This creates periodic nodes and antinodes in the potential, enabling precise control of charges and particles[43,44]. If the RF excitations applied to the left $IDT_1$ and the right $IDT_2$ are considered as two analog input signals, then the SSAW can be utilized to perform addition and subtraction operations on these analog signals, the total combined potential acts upon the FET at the center of the device can be represented as follows:

$$V_{total} = U_1 \cos(\omega t - kx) + U_2 \cos(\omega t + kx + \varphi) \qquad (2)$$

$$= U_t \cos(\omega t - kx - \varphi/2) + U_s \cos(kx + \varphi/2)\cos(\omega t) \qquad (3)$$

In this expression, $U_t = U_1 - U_2$, $U_s = U_1 + U_2$, $U_1$ and $U_2$ represent the amplitude of the piezoelectric potential induced by the left and right SAWs, k is the wave vector of SAWs, ω is the frequency of the SAWs. φ is the relative phase difference of both SAWs. In Equation (3), the first term corresponds to the potential of a traveling wave, while the second term represents the potential of a standing wave, analogous to a traveling wave that carries an oscillating standing wave. The characteristic of the maxima and minima in acoustoelectric current arises from the transition from node to anti-node regions.

Upon applying RF excitation of same intensity to both left $IDT_1$ and right $IDT_2$, the acoustoelectric current remains undetectable due to its fundamental reliance on the directional propagation of charges by SAWs. In this scenario, there is no displacement associated with SSAW, since the intensity of the traveling wave $U_t = U_1 - U_2 = 0$. Nonetheless, when $U_1$ is set to 2.5 V and $U_2$ to 0.5 V, voltage oscillations caused by acoustoelectric charges become detectable with an oscilloscope, as depicted in Fig.5b. This occurrence is attributed to the non-zero traveling wave component of the total combined potential at this point. The oscillating standing wave field either strengthens or weakens the potential that influences charge transport at the dynamic SAW potential minima. The observed differences in the magnitude of the positive and negative voltage can be attributed to the oscillatory nature of the standing wave and the respective locations of probes 1 and 2 in relation to the nodes and anti-nodes of the standing wave.

The phase and amplitude of the output signals can be modulated by adjusting the relative phase and intensity of two counter-propagating SAWs. Fig.5c exhibits the phase control with a periodicity of 2π. The modulation is caused by changes in the relative phase of two counter-propagating SAWs, which causes the wave's nodes and anti-nodes to shift in both position and size relative to probes 1 and 2. By adjusting the phases of the two counter-propagating SAWs, we achieve addition and subtraction operations on the analog signals. When the phase difference is 0, the coherent enhancement maximizes the output signal intensity, facilitating the addition operation; conversely, when the phases differ by π, the signals cancel each other out, resulting in minimal output signal intensity, thus accomplishing the subtraction operation. Fig.5d presents two acoustoelectric current traces, where the excitation voltage for right $IDT_2$ is attenuated by approximately 14 dB and 8 dB compared to left $IDT_1$. The DC measurement of acoustoelectric current exhibits sinusoidal oscillations that correspond to the relative phase shift. As the ratio of $U_2/U_1$ increases, the overall potential becomes more sensitive to changes in phase. Consequently, this results in a larger acoustic current oscillation amplitude.

The acoustoelectric platform modulates the output signal by adjusting the relative phase and intensity ratio of two counter-propagating SAWs, thereby facilitating the addition and subtraction of analog signal transportation.21

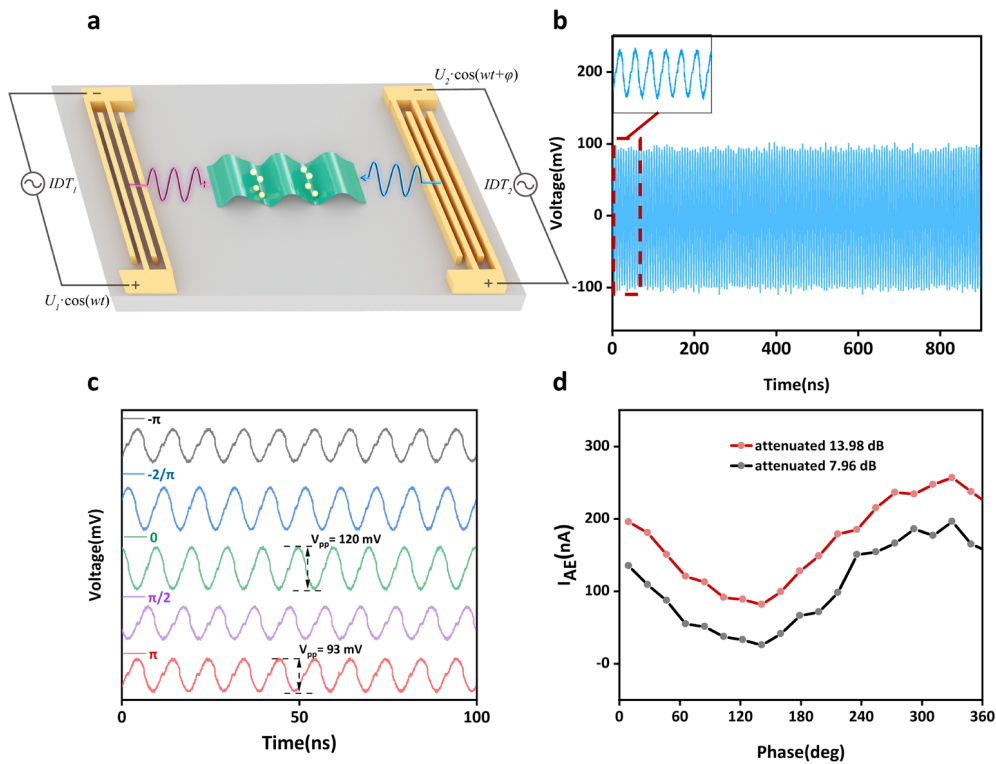

**Fig.5 | Manipulation of carriers through standing surface acoustic wave (SSAW). a,** Schematic of carrier manipulation using SSAW. The pink line represents the incident wave $U_1 \cos(\omega t)$ excited by the left $IDT_1$, the blue line represents the incident wave excited by the right $IDT_2$, $U_2 \cos(\omega t + \varphi)$. **b,** The alterations in voltage across the terminals of the device controlled by SSAW, measured via oscilloscope over an extended duration. Insert of b: the results of measurements taken within a short time span. **c,** Dependence of acoustoelectric current induced voltage across the terminals of the device, on the relative phase of both SAW beams, measured using AC test. **d,** Dependence of the acoustoelectric current on the relative phase of both SAW beams, measured using DC test. The power applied to the first IDT was $V_{pp}$= 5 V. For the second IDT, the source was attenuated by 13.98 and 7.96 dB, respectively.

## Conclusions

We have constructed a hybrid acoustoelectric platform by integrating $WSe_2$ FET and SAW devices. We first investigate the bipolar transport behavior of $WSe_2$ FET under Ti contact, which result in high carrier concentrations of both electrons and holes. Then, with the measured acoustic current almost an order of magnitude larger than prior experimental results[16,45], we confirm the directed transport of charge carriers driven by SAWs. The platform's temporal response is tested using an oscilloscope, revealing a delay time of 2.4 microseconds per centimeter, indicating its suitability for constructing efficient SAW delay line devices. In processing AC signals, the device shows that it can reproduce and regulate the intensity of AC signals. It also has a high dynamic range and can achieve a high on-off ratio by dual modulating the electric gate and SAW intensity. Experiments using standing waves have also allowed for the addition and subtraction of AC signals, indicating that it has the capacity of processing complicated AC signals. In conclusion, this study explores the potential use of 2D material-based acoustic charge transport (ACT) devices for analog AC signal processing. It suggests that the development of ACT devices that are miniaturized, integrable, high-frequency, and wide-bandwidth is the way forward.

## Methods

### Sample fabrication

The IDTs design wavelength was set at $\lambda_{SAW} = 25$ μm, which corresponds to a design frequency of 160 MHz, with a 1:1 duty cycle and an acoustic aperture of $W = 6.2$ μm. Positioned at opposite ends of the substrate, forming a delay line configuration with a length of 2.5 mm, allowing for the generation and detection of SAWs that propagate at a velocity of 3980 m/s in opposite directions across the chip. Few-layer $WSe_2$, h-BN and Graphene flakes were mechanically exfoliated onto a target $LiNbO_3$ substrate by scotch tape in order. Their thickness was measured using atomic force microscopy (Dimension Icon from BRUKER), and the heterostructures were vacuum annealed at 200 °C for two hours. The device fabrication was accomplished using standard photolithography, followed by metal deposition and lift-off techniques. Ti and Au were deposited via evaporation at a rate of 1 Å s$^{-1}$.

### Measurement techniques

The scattering parameters $S_{11}$ (reflection) and $S_{21}$ (transmission and insertion loss) of the IDTs were assessed using a vector network analyzer (Rohde & Schwarz ZNA67) to measure the scattering parameters of the RF network. DC acoustoelectric current was measured using a Keithley 4200A-SCS source meter unit (SMU). The gate voltage was applied by this unit as well. AC time-domain testing was conducted using an oscilloscope (Keithley MSO44).

# Supplementary Information

**Table of Contents**

## Supplementary Note 1
## Raman spectroscopy testing

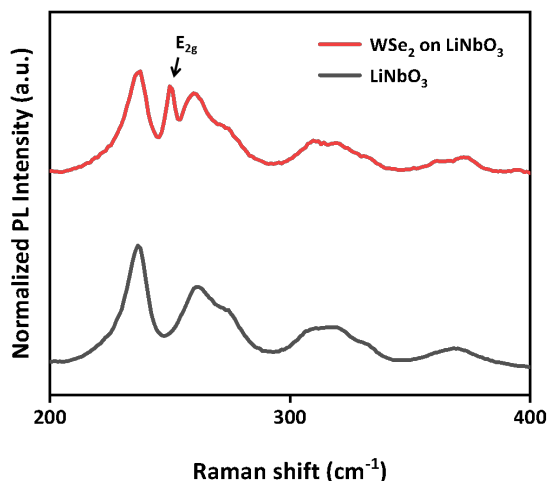

**Supplementary Figure 1. The results of the Raman optical testing.** The Raman spectrum of the LiNbO$_3$ substrate and the WSe$_2$ on the LiNbO$_3$ substrate.

In Supplementary Figure 1, the black line represents the Raman spectrum of the LiNbO$_3$ substrate, while the red line depicts the Raman spectrum obtained from the transfer of WSe$_2$ onto the LiNbO$_3$ substrate via a dry transfer method. Due to the overlap of the Raman peaks, we were only able to observe a single E$_{2g}$ characteristic peak of WSe$_2$[1]. PL testing result is displayed in Figure 1(f), where we observed the A peak corresponding to the direct bandgap and the I peak associated with the indirect bandgap. Notably, the intensity of the I peak was significantly greater than that of the A peak. The bandgap structure of WSe$_2$ varies with changes in layer thickness, and the test results indicate that the number of layers of WSe$_2$ is approximately 12 layers.[2]

## Supplementary Note 2
## Simulation of the total displacement and electric potential of SAW

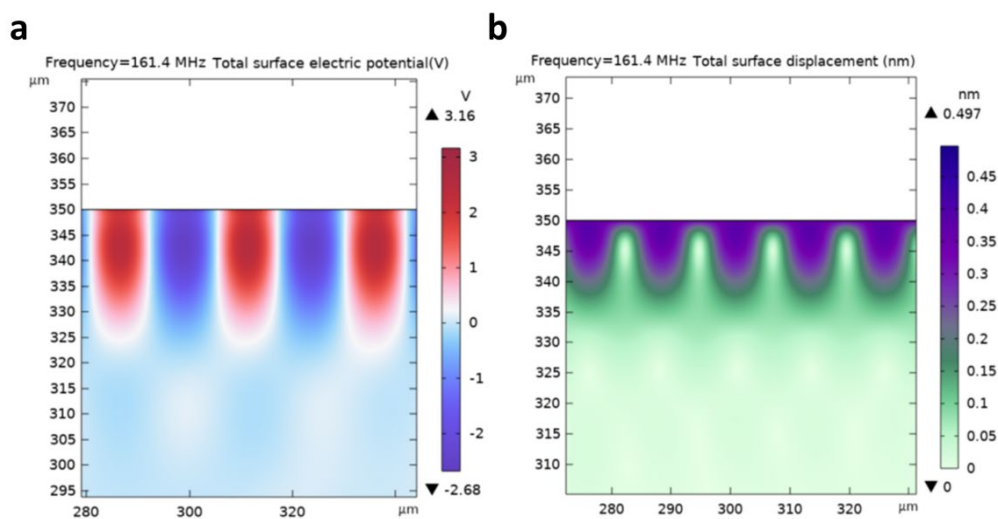

**Supplementary Figure 2. Potential generated by SAW and the resulting total displacement.** (a) Simulation results of the electric potential generated by SAW on the LiNbO$_3$ surface. (b) Simulation results of the displacement generated by SAW on the LiNbO$_3$ surface.

To drive the directional transport of carriers in two-dimensional materials, it is essential for SAWs to provide a sufficiently large potential to confine electrons and holes. Utilizing COMSOL simulation, we determined that on a 128°Y-X cut LiNbO$_3$ substrate with IDTs spacing of 6.25 μm, the simulated resonant center frequency is 161.4 MHz. When an RF excitation power of 18 dBm is applied to the IDTs, the potential generated by the surface acoustic wave on the LiNbO$_3$ substrate surface is shown in Supplementary Figure 2(a), and the resulting total displacement is illustrated in Supplementary Figure 2(b). For an ACT device to work, the peak-to-peak amplitude of the potential distribution must be large enough to prevent electrons from jumping over the barrier from one potential well to the next. At absolute temperature T, the charge carriers in a packet have a range of kinetic energies given by the Fermi-Dirac distribution:

$$f(E) = \frac{1}{1 + e^{\frac{|E-E_f|}{kT}}}$$

This formula provides the probability that an available energy state at E will be occupied. $E_f$ is the Fermi level, and most electrons and holes in a charge packet have energies close to the conduction band and valence band level. However, the Fermi-Dirac distribution indicates that the probability of electrons and holes having higher energy levels is not zero and increases with temperature T. Consequently, there will always be some electrons and holes in a packet that have enough energy to jump over the barrier to an adjacent well. Nevertheless, if the barrier height is set sufficiently high, approximately 10kT (≈0.259 V at room temperature) or more, the probability of such occurrences will be significantly reduced. Utilizing Hoskins' theory[3], this approach treats electron and holes diffusion over the crests of the potential wave as a small perturbation of the diffusionless problem. From the simulation results in Supplementary Figure 2(a), it is evident that the barrier height created by the SAW's electric potential significantly exceeds 10kT, thus it can be inferred that the carriers are being carried along by SAW for directional transport, rendering diffusion negligible.

**Supplementary Note 3**
**Theoretical calculation of the acousto-electric current magnitude**
Here, we utilize the research by Weinreich et al. on the movement of electrons and holes in semiconductors under the influence of SAW as our theoretical foundation[4-7]. In this model, the relaxation time τ for the redistribution of charged particles under the influence of SAW to reach equilibrium is neglected.
In his assumption, the potential energy carried by the SAW acts on the particles, resulting in the generation of acousto-electric current. In brief, the particle current j in the macroscopic system comprises the drag term and the diffusion term, which can be

expressed as follows:

$$j = D(\frac{F}{KT} - DV) \tag{S1a}$$

n represents the particle density, V represents the potential energy of the particles, F represents the force on the particles, KT stands for thermal energy, and D is the diffusion coefficient. The plane wave in the semiconductor has the following form:

$$\Phi = \Phi_0 e^{ik(x-ct)} \tag{S2a}$$

where $\Phi_0$ is a constant, k is the wave vector, and c is the velocity of the acoustic wave moving in the x-direction. The energy exchange between particles and the electric field can be expressed as:

$$U_I = \sum_i q_i \Phi_i \tag{S3a}$$

If a longitudinal electric field is applied to give the charge carriers a drift velocity uc in the x-direction, the electron and hole acousto-electric current densities, $j_n$ and $j_p$ respectively, are given by:

$$j_{n(p)} = n(p)uc + \frac{n(p)D_{n(p)}}{kT}\frac{\partial}{\partial x}(q_{n(p)}\Phi \pm e\gamma_k\Phi) - D_{n(p)}\frac{\partial n(p)}{\partial x} \tag{S4a}$$

In the equation, $n$ and $p$ represent the concentrations of electrons and holes, respectively. $q_n$ and $q_p$ correspond to the charge quantities of electrons and holes, respectively. $kT$ signifies thermal energy, while $\gamma_k \Phi$ refers to the induced electrostatic potential. The continuity equation takes the form:

$$\frac{\partial n}{\partial t} + \frac{\partial j_n}{\partial x} + \frac{1}{\tau}[n - n_0(1 + \frac{q}{1+m}\frac{\Phi}{kT})] = 0 \tag{S5a}$$

$$\frac{\partial p}{\partial t} + \frac{\partial j_p}{\partial x} + \frac{1}{\tau}[p - p_0(1 + \frac{qm}{1+m}\frac{\Phi}{kT})] = 0 \tag{S5b}$$

where $m = n_0/p_0$ is the ratio of equilibrium concentrations of electrons and holes, $q = q_n + q_p$, and τ is the lifetime. Assuming $n - n_0 \ll n_0$ and $p - p_0 \ll p_0$, and let

$$n = n_0 + n_1 e^{ik(x-ct)} \tag{S6a}$$
$$p = p_0 + p_1 e^{ik(x-ct)} \tag{S6b}$$

By synthesizing Eqn. S4, S5, S6, and considering charge conservation along with the impact of charge relaxation time, we introduce a parameter $\gamma_\tau$ related to the relaxation time τ. Ultimately, we derive the expression for the acousto-electric current as follows:

$$I_{AE} = -\frac{1}{2}\frac{en_0c}{(1+m)^2}(\frac{q\gamma_\tau\varphi_0}{KT})^2\frac{(1-m)-u(1+m/b)}{1+(\beta_0-u)^2[(1-m)/(1+m)]^2(\omega_0/\omega)^2} \tag{S7}$$

$b$ is the ratio of the electron and hole mobilities, $\beta_0$ defined as $(1 - mb)/(1 - m)$.

**Supplementary Note 4**
**Variations in the output and transfer characteristics of WSe$_2$ FET with and without SAW.**

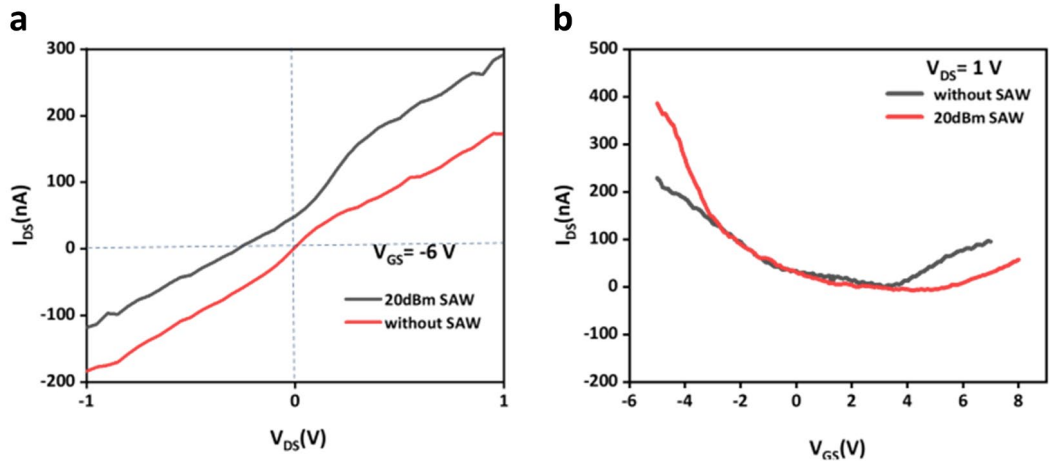

**Supplementary Figure 3. Output and transfer characteristics of WSe$_2$ FET with and without SAW.** (a) Output curves of the WSe$_2$ FET when V$_{GS}$ = -6V with (black curve) and without (red curve) the SAW. (b) Transfer curves of the WSe$_2$ FET when I$_{DS}$ was driven by the electrical field only (black curve) and the SAW as well (red curve), V$_{DS}$=1V.

**Supplementary Note 5**
**Changes in SAW potential before and after passing through the 2D FET .**

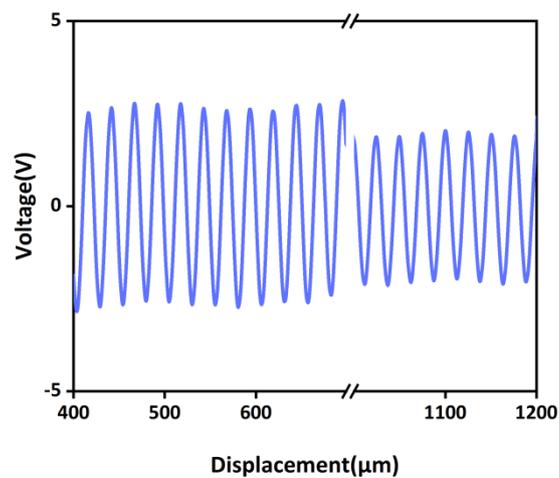

**Supplementary Figure 4. The COMSOL simulation results of the SAW potential before and after passing through the FET region of the two-dimensional materials.** The SAW potential experiences a degree of attenuation after passing through the FET region composed of graphene, h-BN, and WSe$_2$. The conductivity of the graphene significantly decreases due to doping and exposure to air. Since the exact conductivity value of the FET region is unknown, we incorporated a semiconductor material with a certain conductivity along the propagation path of the SAW in the simulation to approximate the attenuation of the SAW potential.

**Supplementary Note 6**

**Pulse responses of the device.**

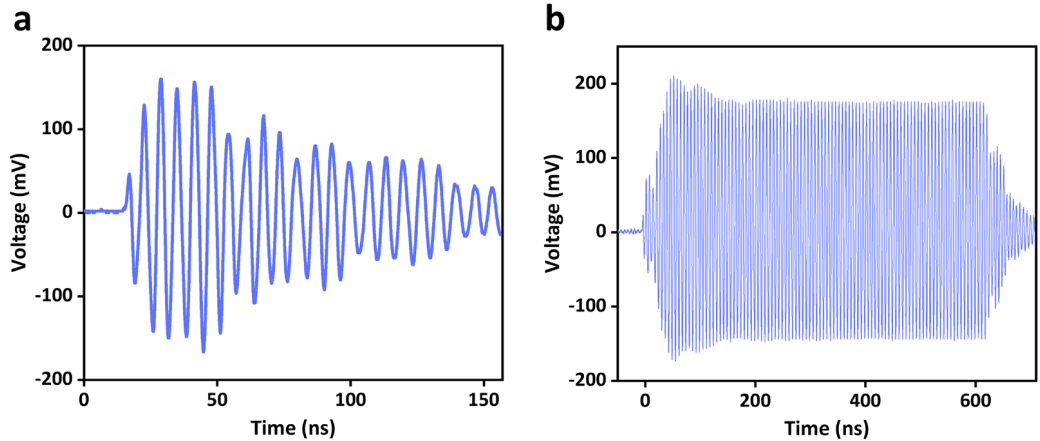

**Supplementary Figure 5.** The device's response to (a) five pulse sine waves and (b) one hundred pulse sine waves.